\documentclass[12pt]{article} 
\usepackage{hyperref}
 \usepackage{amsmath}
 \usepackage{amssymb}
 \usepackage{graphicx}
 \usepackage{color}
 \newlength{\abstractwidth}
 \newcommand{\be}{\begin{equation}}
 \newcommand{\bea}{\begin{eqnarray}}
 \newcommand{\eea}{\end{eqnarray}}
 \newcommand{\beq}{\begin{equation}}
 \newcommand{\ee}{\end{equation}}
 \newcommand{\eeq}{\end{equation}}

\def\32{{3 \over 2 } }

 \def\ba{\begin{eqnarray}}
 \def\ea{\end{eqnarray}}

 \def\simleq{\; \raise0.3ex\hbox{$<$\kern-0.75em
 \raise-1.1ex\hbox{$\sim$}}\; }
 \def\simgeq{\; \raise0.3ex\hbox{$>$\kern-0.75em
 \raise-1.1ex\hbox{$\sim$}}\; }

\def\nref#1{(\ref{#1})}

\begin{document}

\begin{titlepage}
\bigskip
\bigskip
\bigskip 
\bigskip
\bigskip

\begin{center}
\bigskip
{\Large \bf { 
Primordial gravity wave background anisotropies 
 }}
 
\bigskip
\bigskip
\bigskip
\bigskip
\bigskip
\bigskip
\bigskip

Vasyl Alba$^1$ and Juan Maldacena$^2$

\bigskip 
\bigskip

$^1$ { \it 
Department of Physics, Princeton University, Princeton, NJ, USA\\ }
\smallskip 

 $^2$ { \it Institute for Advanced Study, 
 Princeton, NJ, USA\\ }
 
\bigskip
\bigskip
\end{center}

\bigskip
\bigskip
\begin{abstract}
We consider the primordial gravity wave background produced by inflation. We 
compute the small anisotropy produced by the primordial scalar fluctuations.
\medskip
\end{abstract}
\bigskip
\bigskip
\bigskip
\bigskip 
\end{titlepage}

\section{Introduction} 
 
The theory of inflation predicts a stochastic background of gravity waves, which were 
produced by quantum effects \cite{Starobinsky:1979ty,Starobinsky:1980te,Guth:1980zm,Linde:1981mu,Albrecht:1982wi,Mukhanov:1981xt,Hawking:1982cz,Guth:1982ec,Starobinsky:1982ee,Bardeen:1983qw}. 
 To leading order, this is a statistically homogeneous and isotropic background. 
However, at higher orders we expect some influence from the scalar fluctuations. In the same way that the cosmic microwave radiation is not isotropic, but has small 
anisotropies of order $10^{-5}$, the gravity wave background is also not 
isotropic but has small anisotropies of the same order, also due to the scalar fluctuations. 
At relatively large angles, which correspond to scalar fluctuations entering the horizon during matter 
domination, the main effect is due to the Sachs-Wolfe effect \cite{Sachs:1967er}. 
Namely we start locally with 
the same spectrum of gravity waves, which are then red or blue shifted by an amount proportional to the 
amplitude of the scalar fluctuations. The overall observable effect then depends on the frequency dependence
of the spectrum of the gravity waves, which is determined by the properties of the universe when 
the gravity wave mode crosses the horizon, both exiting and reentering.

Of course, given that we have not yet measured the leading order gravity waves, the 
effects that we discuss in this paper are not going to be measured in the near future.
However, it is a theoretically interesting effect, and we trust in
the ingenuity of current and future experimentalists!

This paper is organized as follows. We first give a quick review of the leading order gravity wave background. 
Then, we discuss its anisotropies, focusing on relatively large angles which correspond to scalar 
fluctuations that entered the horizon after matter domination. We also focus on gravity waves that entered the horizon 
in the radiation dominated era, which would correspond to wavelengths that could be measured directly 
by gravity wave detectors. 
 
\section{Gravity waves from inflation at leading order}

\begin{figure}[h]
\begin{picture}(400,55)
\put(0,-38){\line(0,1){37}}
\put(118,-38){\line(0,1){37}}
\put(0,-1){\line(1,0){118}}
\put(-5,30){$\frac{\exp(ik\eta)}{a(\eta)}$~,}
\put(39,-22){Inflation}
\put(38,30){~~$(\eta)^\mu H^{(2)}_{-\mu}(-k\eta)$~,}
\put(60,25){\vector(0,-1){25}}
\put(115,30){~~~~constant~,}
\put(194,-22){$p=w\rho$}
\put(175,30){~~~~$\eta^{-\nu+1/2}J_{\nu-1/2}(k\eta)$~,}
\put(209,25){\vector(0,-1){25}}
\put(330,30){$\cos \left(k\eta +\phi_0 \right)/a(\eta)$~}
\put(306,-38){\line(0,1){37}}
\put(395,-38){\line(0,1){37}}
\put(306,-1){\line(1,0){89}}
\put(332,-16){Matter}
\put(323,-30){domination}
\thicklines
\linethickness{0.3mm}
\put(-10,-120){\line(1,0){470}}
\put(-10,55){\line(1,0){470}}
\put(-10,-120){\line(0,1){175}}
\put(460,-120){\line(0,1){175}}
\linethickness{0.5mm}
\put(0,-38){\line(1,0){125}}
\put(170,-38){\line(1,0){70}}
\multiput(125,-38)(6,0){8}{\line(1,0){2}}
\multiput(240,-38)(6,0){9}{\line(1,0){2}}
\put(60,-43){\line(0,1){10}}
\put(118,-43){\line(0,1){10}}
\put(209,-43){\line(0,1){10}}
\put(320,-43){\line(0,1){10}}
\put(395,-43){\line(0,1){10}}
\put(290,-38){\line(1,0){165}}
\end{picture}
\includegraphics[width=\textwidth]{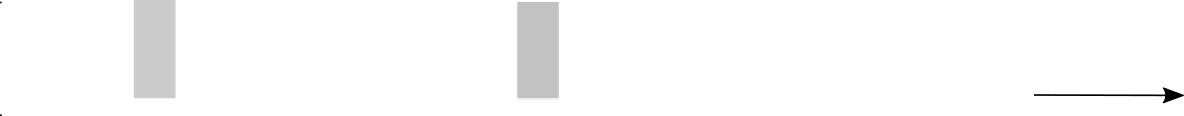}
\begin{picture}(400,65)
\put(55,55){$\eta_{*,k}$}
\put(40,35){Horizon}
\put(40,24){crossing}
\put(40,12){during}
\put(40,0){Inflation}
\put(190,55){$\eta_{\times,k}=\nu/k$}
\put(190,35){Horizon}
\put(190,24){reentry}
\put(190,12){of $h(k)$}
\put(105,35){End of}
\put(105,24){Inflation}
\put(313,55){$2/k_L$}
\put(305,35){Horizon}
\put(305,24){reentry}
\put(305,12){of $\zeta(k_L)$}
\put(385,35){Now}
\put(438,55){$\eta$}
\end{picture}
\caption{Summary of the evolution of a single mode during the history of the universe. We need to know the equation of state only in the shaded regions. 
The first shaded region occurs during slow-roll inflation. The second shaded region takes place during an
 epoch with equation of state $p=w\rho$. Both in the far past and present time the WKB approximation is valid. Outside the horizon the solution has constant amplitude. 
 Long wavelength scalar fluctuation modes entered horizon during the matter dominated era that is assumed to last until now. One the top line we have written the form of the solution in each region. }
\label{History}
\end{figure}
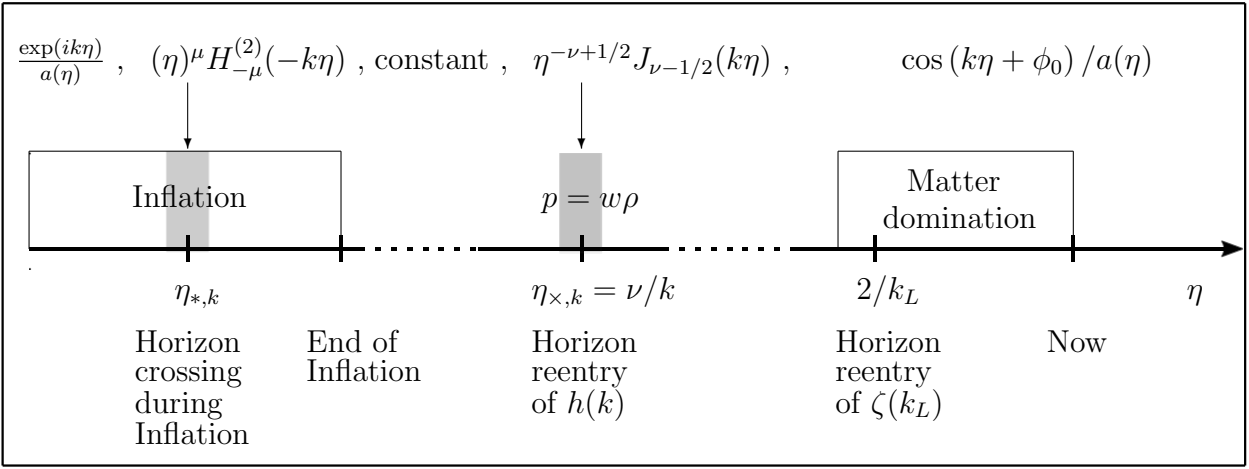

We can view the leading order effect as the result of a Bogoliubov transformation between the 
vacuum in the very early inflationary period and the vacuum today \cite{Starobinsky:1979ty}. In other words, expanding the
Einstein action to quadratic order we get an action for each of the two polarization components of
the gravity waves. Each polarization component obeys an equation equal to that of a minimally 
coupled scalar field in the spatially uniform time dependent cosmological solution. This background
metric can be written as $ds^2 = a(\eta)^2 ( - d \eta^2 + d \vec x^{\,2} ) $, where $\eta $ is conformal time. 
Due to translation symmetry, we can focus on one comoving momentum mode $\vec k$ at a time. 
Each of these modes undergoes the following history. It starts its life well inside the horizon in the 
adiabatic vacuum. It then exits the horizon during inflation at some 
 time $\eta_{*,k} $, where $\eta_{*,k} k =1$. It then remains
with constant (time independent) amplitude until it reenters the horizon. When the mode is well inside
the horizon it is oscillating and redshifting as any other massless particle. See figure \ref{History}. 
The solution well inside the horizon is given by the WKB approximation. In order to match
 the solution well 
inside to the one well outside the horizon, we need to solve the equation 
 during horizon crossing. During horizon crossing we can approximate
the evolution of the background in terms of a fluid with a constant $w$, where 
$p = w \rho$, so that $a(\eta) \propto \eta^{ 2 \over 1 +3 w } $. Then the solution becomes a combination of Bessel functions. When the mode 
is exiting the horizon during inflation, the appropriate solution is proportional to 
$ \eta^{\mu} H^{(2)}_{- \mu} (-k \eta) $, with $\mu = { 3 \over 2 } + \epsilon$, where $\epsilon $ is the 
standard slow roll parameter ($ \epsilon =- \dot H/H^2 $) and $k = |\vec k |$. 
Similarly, when the mode reenters the horizon when the universe is dominated by a fluid with 
an equation of state with fixed $w$, then the solution is proportional to $\eta^{ -\nu + {1 \over 2} } J_{\nu - {1\over 2 } }( k \eta ) $, 
with $\nu = 2/(1 + 3 w) $. 
We will call the time of horizon reentry, $\eta_{\times , k}$, defined by 
setting $ k = a(\eta_{\times, k} ) H(\eta_{\times , k}) $ or $k \eta_{\times , k} = \nu $.
 
Matching the solutions we find that the normalized solution has a late time behavior of the form 
\begin{equation} \label{WaveAmplitude}
h(k,\eta)=\frac{H(\eta_{*,k_0} )}{\sqrt{2} k^{\frac{3}{2}} M_{pl}}\left(\frac{k_0}{k}\right)^{ \epsilon + \nu} \frac{ \Gamma \left( \nu+\frac{1}{2}\right)}{\sqrt{\pi} }\left(\frac{2}{\nu}\right)^\nu \frac{a(\eta_{\times,k_0}) }{a(\eta)}\cos \left(k\eta +\phi_0 \right),
 \end{equation}
Here $k_0$ is a reference comoving momentum\footnote{In other words, if we are interested in a certain range of values of $k$, we can choose $k_0$ to be somewhere
in that range. There is no dependence on $k_0$ due to the dependence on $\eta_{*,k_0}$ and $\eta_{\times, k_0}$. }. 
The factor of $ \left(\frac{k_0}{k}\right)^{ \epsilon + \nu} $ arises because, as we move $k$ away from $k_0$, we are crossing the horizon earlier or later. Here 
$\phi_0$ is the initial phase of the wave, which is $-\frac{\pi \nu}{2}$, but is unmeasurable for the
gravity waves we consider. The redshift after horizon reentry is encoded in the factor $1/a(\eta)$ in 
(\ref{WaveAmplitude}). 
 
The final density of gravity waves is proportional to the square of (\ref{WaveAmplitude}). We can 
express it in terms of the energy density or number density as follows 
\bea
 d\rho_g &=&2\omega \frac{\omega^2 d\omega}{(2 \pi)^3}d\Omega \left<N_{\vec k}\right>,
~~~~{\rm with} ~~~ \label{Spectrum}
\\
\left< N_{\vec k }\right>&=&{ H(\eta_{* , k})^2 \over H(\eta_{\times, k })^2 } C_\nu=
{ H(\eta_{* , k_0})^2 \over H(\eta_{\times, k_0 })^2 } \left(\frac{k_0}{k}\right)^{2(\nu+1+\epsilon)} C_\nu \notag 
\\
C_\nu &\equiv& { \Gamma(\nu +{1 \over 2 } )^2 \over 4 \pi } \left( { 2 \over \nu }\right)^{2 \nu} 
\notag \eea
where $\left< N_{ \vec k }\right>$ is a density of states as a function of $k$ and $\omega = k/a$ is the physical energy of the waves. The factor of two comes 
from the two polarization states. It is useful to think in terms of $N_k$ because this is conserved 
after horizon reentry. Notice that $N_k$ is essentially given by the ratio of the Hubble constants at 
horizon crossing. We have also made more explicit the form of the spectrum by defining an arbitrary 
reference momentum $k_0$. 
 In the case of photons we would get the usual thermal distribution for $N_k$.
 We see that the spectrum is determined by the equation of state at the horizon crossing times. More generally, the spectrum of gravity waves encodes the
 whole expansion history of the universe \cite{Boyle:2005se,Nakayama:2008wy,Zhang:2006mja,Kuroyanagi:2011fy,Jinno:2013xqa}.

\section{Anisotropy}

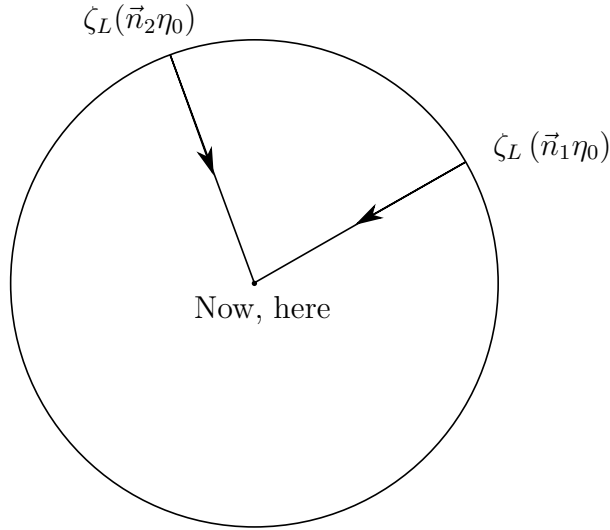
\begin{figure}[h]
\includegraphics[width=\textwidth]{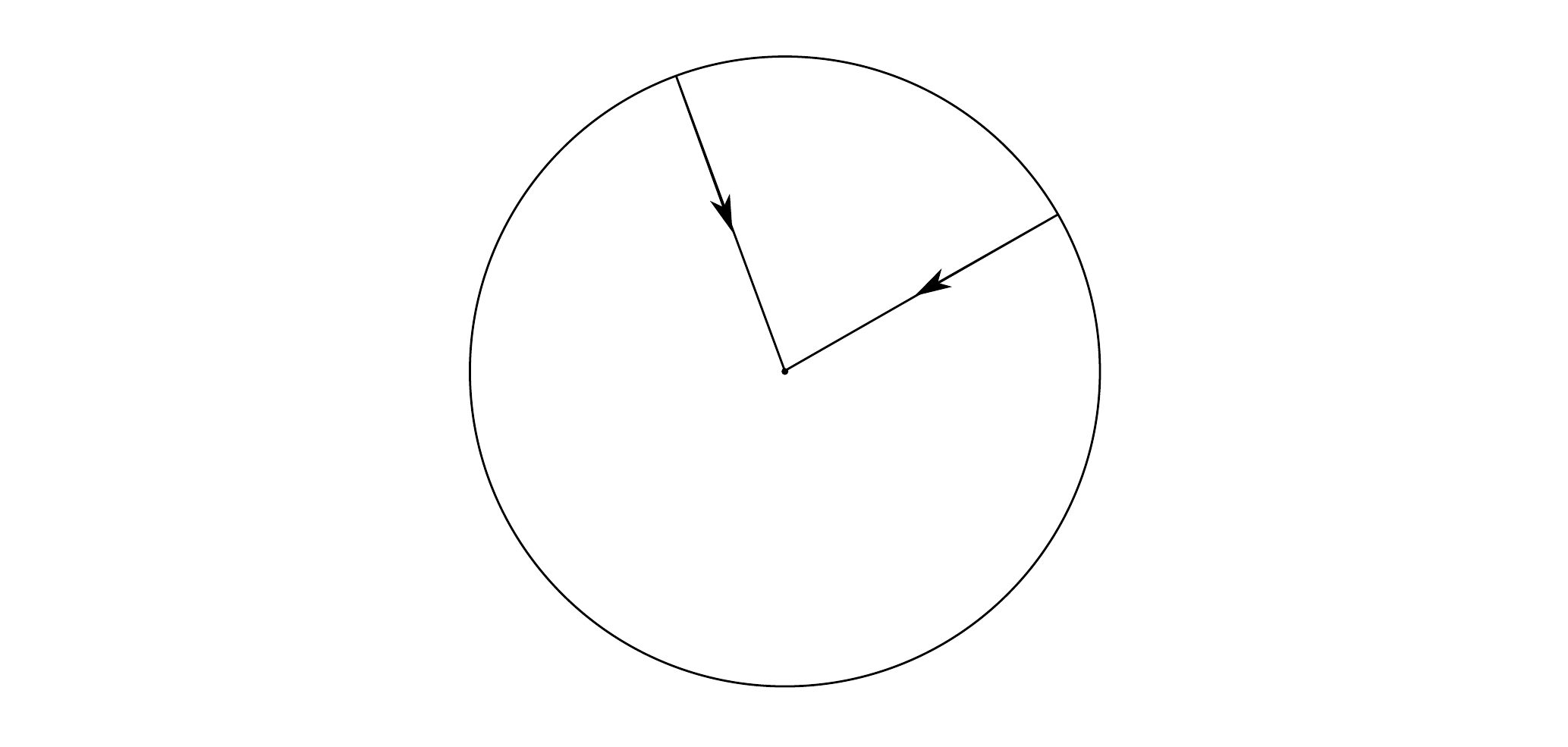} 
\caption{ Gravitons that are coming from different directions have different redshift, because values of a scalar perturbation $\zeta_L$ are different at different emission points. }
\begin{picture}(460,-10)
\put(207,150){Now, here}
\put(320,212){$\zeta_L\left(\vec n_1\eta_0\right)$}
\put(165,260){$\zeta_L(\vec n_2\eta_0)$}
\end{picture}
\label{anisotropy}
\end{figure}

We are interested in the anisotropy of the gravity waves that comes from the interaction with the scalar fluctuations. We will concentrate on the anisotropy at 
relatively large angular scales, which is produced by long wavelength scalar fluctuations. More precisely, we consider a scalar fluctuation mode with a momentum 
$k_L$ which reenters the horizon during the matter dominated era. In order to find the effect on the gravity waves, it is important to solve the equation of motion for
the mode $\zeta_{k_L}$ after it reenters the horizon. We then consider the propagation of the gravity waves through that perturbed universe. This computation 
is essentially identical to one done for photons, which is the so called Sachs-Wolfe effect \cite{Sachs:1967er}. 
We can work in the gauge where the matter density is constant so that the fluctuation is purely on the scale factor of the spatial part of the metric of the surfaces with 
constant density. In that gauge $\zeta_{k_L}$ continues to be constant after crossing the horizon, but the metric develops a $g_{0i}$ component that should be
computed by first considering $w>0$ and taking $w \to 0$ at the end of the computation \cite{Boubekeur:2008kn}\footnote{
See equations (35)-(38) in \cite{Boubekeur:2008kn}. We thank M. Zaldarriaga for sharing some notes on this with us. }. 
Then the perturbed metric has the form $ds^2 = a^2 [ -d\eta^2 + (1 + 2 \zeta) d\vec x d\vec x - { 4 \over 5 a H } \partial_i \zeta d\eta dx^i ] $. 
Solving the geodesic equation on this perturbed metric we find that the extra redshift for a massless particle emitted at an early time and observed today 
is 
\be 
{ \delta \omega \over \omega } = { 1 \over 5} \left[ \zeta_{\rm today} - \zeta_{\rm emitted} \right] 
\ee
where $\zeta_{\rm today }$ is the value at the location of the observations today, while $\zeta_{\rm emitted} $ is the value of $\zeta$ at the location where 
the massless particle was emitted, when the mode $\zeta_{k_L}$ was outside the horizon (so that we can neglect the ``doppler term''\footnote{We have also ignored the doppler term today. It is trivial to put it back in.}). 
Since $\zeta$ is time independent, $\zeta_{\rm emitted} = \zeta( \vec x_e = \vec n \eta_0 )$ where $\eta_0$ is the present value of the conformal time and 
$\vec n$ is the direction that the massless particle is coming from\footnote{Note that $\vec x_e = \vec n( \eta_0 - \eta_e) \sim \vec n \eta_0$, and $\vec x_{\rm today, ~here} = \vec 0 $.}.
 We then assume that at the emission time the local physics is completely independent of 
the long scalar mode, which is well outside the horizon. As usual, 
when we compare the energy coming from 
different directions $\zeta_{\rm today} $ cancels out, see figure \ref{anisotropy}. 

The discussion so far is identical to the discussion of the Sachs Wolfe effect for the CMB. The only difference is that the spectrum of gravitons 
 is not thermal, but it is instead given by a power law distribution $ (\omega_0/\omega)^{ 2 (\nu + 1 + \epsilon) }$, \nref{Spectrum}. Instead of $\left(\delta T/T\right)_{SW} = - \zeta/5$, we simply vary $\delta \omega_0/\omega_0 = -\zeta/5$. More explicitly, if we look at the spectrum of gravity waves as a function of frequency $\omega$ coming from the direction $\vec n$, we obtain 
\be 
d\rho(\omega, \vec n) = d\rho (\omega) \left \{ 1 + { 2 \over 5} ( \nu + 1 + \epsilon ) \big[ \zeta_{\rm today} - \zeta_L( \vec n \, \eta_0 ) \big] \right\} 
\ee
were $d\rho(\omega)$ is the isotropic part of the spectrum, given in (\ref{Spectrum}). This result can 
be derived also using cosmological perturbation theory, performing the computation of the three point function as in \cite{Maldacena:2002vr} and following the evolution to the present, the details will be 
presented separately \cite{Alba}. 
Here we have neglected the integrated Sachs Wolfe effect which is due to the cosmological constant. This can be taken into account, as in the case of photons. 
Since the same effect is giving rise to fluctuations in the CMB and in the gravity wave spectrum, we can also write the formula as 
\be \label{final}
d\rho(\omega, \vec n) = d\rho (\omega) \left \{ 1 - \left( { \delta T \over T }\right)_{SW + ISW} \times k\partial_{k} \log \langle N_k \rangle \right\} 
\ee
where the subindex indicates the contribution from the Sachs-Wolfe and the integrated Sachs Wolfe effect. 
 The discussion so far has not included the damping effect of the neutrinos \cite{Weinberg:2003ur}. 
This is expected to be a local effect which will not depend on the long mode. For that reason the final formula as written in (\ref{final}) would also be correct
if one inserts the full $\langle N_k\rangle $ expression that includes the effects of the neutrinos. 
We can view this equation as a consistency condition for a single field inflation. Usually, the consistency condition is discussed for the wavefunction of the universe
outside the horizon \cite{Maldacena:2002vr,Creminelli:2004yq} (see also \cite{Creminelli:2011sq}), but as emphasized in \cite{Pajer:2013ana} the physical content of that condition is that 
a local observer cannot notice the long fluctuation. It is a manifestation of the equivalence principle. Furthermore, the final answer could be viewed as arising from 
a projection effect due to the propagation of the massless particles through the perturbed universe. 
Furthermore, since the whole effect comes from late time projection effects, the final formula
\nref{final} is valid 
also for gravity waves that are generated by any process that happened before the long mode crossed the horizon, such 
as phase transitions in the early universe, for example\footnote{We thank M. Zadarriaga for making this observation.}.

Any deviation away from this expression (\ref{final}) would be evidence of a second
field which would be affecting the relative densities of gravity waves relative to everything else. We think that this would occur in the curvaton models
\cite{Enqvist:2001zp,Lyth:2001nq,Moroi:2001ct}. 

Since the fluctuations are very small, of order $10^{-5}$, these are rather difficult to measure. In addition, we should remember that we are dealing with a 
stochastic background, so that the observed gravity wave over a small number of cycles is fairly random, and the statement in (\ref{final}) is about the deviation in
the variance of that random variable. This means that we need to measure this random variable many times. For the case of random waves, we can view each 
cycle of the wave as one instance of the random distribution. This means that to reach this accuracy we need to observe of order $10^{10}$ cycles in each angular
direction. For waves a frequency of $f$[Hz] we need about $300/f$ years.

{ \bf Acknowledgments} 

We thank V. Mukhanov and specially M. Zaldarriaga for discussions. 

The work of VA was supported by the National Science Foundation under Grant No. PHY-1314198.

\end{document}